\documentclass{article}
\usepackage[dvips]{graphicx}  
\usepackage{color}

\usepackage[varg]{txfonts}
\usepackage{concmath}

\begin{document}

\newcommand{\rum}{\rule{0.5pt}{0pt}}
\newcommand{\rub}{\rule{1pt}{0pt}}
\newcommand{\rim}{\rule{0.3pt}{0pt}}
\newcommand{\numtimes}{\mbox{\raisebox{1.5pt}{${\scriptscriptstyle \rum\times}$}}}
\newcommand{\numtimess}{\mbox{\raisebox{1.0pt}{${\scriptscriptstyle \rum\times}$}}}
\newcommand{\Boldsq}{\vbox{\hrule height 0.7pt
\hbox{\vrule width 0.7pt \phantom{\footnotesize T}%
\vrule width 0.7pt}\hrule height 0.7pt}}
\newcommand{\two}{$\raise.5ex\hbox{$\scriptstyle 1$}\kern-.1em/
\kern-.15em\lower.25ex\hbox{$\scriptstyle 2$}$}

\renewcommand{\refname}{References}
\renewcommand{\tablename}{\small Table}
\renewcommand{\figurename}{\small Fig.}
\renewcommand{\contentsname}{Contents}

\twocolumn[%
\begin{center}
{\Large\bf 
Observed Gravitational Wave Effects:  Amaldi  1980 Frascati-Rome Classical  Bar 
Detectors, 2013 Perth-London Zener-Diode Quantum Detectors,   Earth Oscillation  Mode Frequencies 
 \rule{0pt}{13pt}}\par

\bigskip
Reginald T. Cahill\\ 

{\small\it School of Chemical and Physical  Sciences, Flinders University,
Adelaide 5001, Australia\rule{0pt}{15pt}\\

Email: Reg.Cahill@flinders.edu.au}
\par

\bigskip

{\small\parbox{11cm}{%
Amaldi {\it et al.} in 1981 reported two key discoveries from the  Frascati and Rome  gravitational wave cryogenic bar detectors: (a) Rome events delayed by within a few seconds to tens of seconds from the Frascati events, and (b) the Frascati  Fourier-analysed data frequency peaks being the same as the  earth  oscillation frequencies from seismology. The time delay effects have been dismissed as being inconsistent with gravitational waves having speed c. However using data from zener diode quantum detectors, from Perth and London, for January 1-3, 2013,   we report the same effects, and in excellent agreement with the Amaldi results. The time delay effects appear to be gravitational wave reverberations, recently observed, and for gravitational wave speeds of some 500km/s, as detected in numerous experiments.  We conclude that the Amaldi {\it et al.} discoveries were very significant.

\rule[0pt]{0pt}{0pt}}}\medskip
\end{center}]{%

\setcounter{section}{0}
\setcounter{equation}{0}
\setcounter{figure}{0}
\setcounter{table}{0}

\section{Introduction}

On the basis of data from the new nanotechnology zener diode quantum gravitational wave detectors \cite{ZenerDiode} it is argued that the wave effects detected in 1980 by Amaldi {\it et al.} \cite{AmaldiA, AmaldiB}, using two cryogenic bar detectors, located in Frascati and Rome,  were genuine gravitational wave effects, together with  earth oscillation effects, although not gravitational waves of the expected form. 

The speed and direction of gravitational waves  have been repeatedly detected using a variety of techniques over the last 125 years, and have a speed of some 500km/s coming from a direction with RA $\sim$  5hrs, Dec $\sim$ 80$^0$. These waves appear to be of galactic origin, and associated with the dynamics of the galaxy and perhaps the local cluster.  This speed is that of the dynamical 3-space, which appears to have a fractal structure, and the significant magnitude waves are turbulence/fractal structure in that flowing space. The detection techniques include gas-mode Michelson interferometers, RF coaxial cable EM speed measurements, RF coaxial-cable - optical fiber  RF/EM speed measurements, EM speed measurements from spacecraft  Earth-flyby Doppler shifts, zener-diode quantum detectors, within Digital Storage Oscilloscopes, and in so-called Random Event Generators (REG); Cahill \cite{CahillNASA,CahillWaves,ZenerDiode}. These zener diode devices have detected correlations between Adelaide and London, and between Perth and London, with travel time delays from  10 to 20 seconds, and with significant reverberation effects  \cite{ZenerDiode,Reverb}.    The speed of some 500km/s has also been observed as a time delay of some 500ns  in table-top zener-diode quantum detectors, separated by 25cm in a S to N direction.  The zener diode gravitational wave quantum detectors operate by the process of the 3-space wave turbulence causing the quantum to classical transition, i.e. spatial localisation, of the electron wave functions  tunnelling through a  10nm quantum barrier, when the diode is operated in reverse bias. The earlier techniques rely on detecting EM radiation anisotropy.

 \begin{figure}[t]
\hspace{1mm}\includegraphics[scale=2.3]{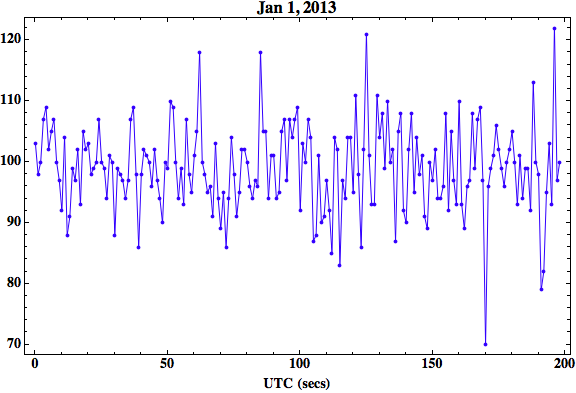}
\vspace{-3mm}	\caption{\small  Perth  zener diode quantum detector (REG) data, for January 1, 2013. The data points are at 1s intervals.     The data shows strong peaks at  5 - 30s intervals, related to the reverberation effect \cite{Reverb}.  This appears to be the time-delay effect detected between  the Frascati and Rome cryogenic gravitational wave bar detectors   \cite{AmaldiA, AmaldiB}.}
\label{fig:Reverb}\end{figure}

\begin{figure*}
 \vspace{-10mm}  \hspace{10mm} \includegraphics[scale=0.95]{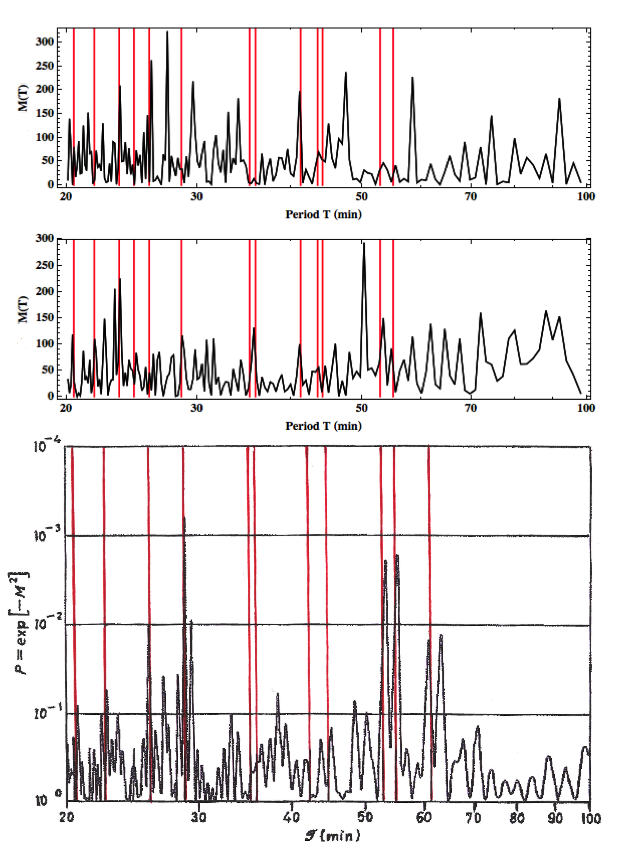} 
   \vspace{-1mm}     \caption{\small  Top: Power spectrum  from  Zener Diode  detector in Perth, Jan 1-3, 2013.  Middle: Power spectrum  from  Zener Diode  detector in London, Jan 1-3, 2013. Bottom: Power spectrum from Frascati  bar detector data, May  6-7, 1980, adapted from Amaldi {\it et al.} \cite{AmaldiB}. Vertical lines (red) show various earth vibration periods, determined by seismology, \cite{Frequencies}. $M(T) = |F(T)|^2$ is the power spectrum, expressed as a function of period $T$, where $F(T)$ is the Fourier transform of the data time series. A  200sec interval of the Perth data  is shown in Fig.\ref{fig:Reverb}. The spectra from all detectors show the same  low frequency peaks, but with differing intensities. The peaks at 53.1 and 54.1 min equal the $_0S_2^{+1}$ and $_0S_2^{-1}$ Earth vibration modes.}
 \label{fig:periods} \end{figure*}

\section{The Amaldi Frascati Rome Gravitational Wave Detectors}

Data was collected with two cryogenic resonant gravitational wave antennas operated simultaneously
in Rome and Frascati. Coincidences were detected with pulses lasting  about 1 second, and travel times differing from one second to twenty seconds ($\pm 0.5$s), with the NW Rome signal delayed relative to the Frascati events.  These events were dismissed as gravitational wave events as the travel times, for the 20km separation, far exceed that expected if one assumes that gravitational waves travel at speed c, predicting travel times   $\sim$0.1ms.  As well frequency analysis of the data revealed strong peaks at frequencies coinciding with known vibration frequencies of the earth, see bottom plot of Fig.\ref{fig:periods}. Amaldi {\it et al.} considered several mechanisms for the detection of such frequencies: (i) various instrumental couplings to the earth vibrations, (ii)  gravitational field variations caused by a terrestrial source.    However the very same results are obtained with the zener diode quantum gravitational wave detectors.

 \section{Zener Diode Detectors} 
 In \cite{ZenerDiode} the discovery of the nanotechnology zener diode detection effect for gravitational waves was reported.  This was established by detecting times delays between wave forms of 10-20 seconds for travel times Adelaide to London, and Perth to London, with that travel time variation following the earth's rotation wrt  the RA and Dec that had been reported in earlier experiments, \cite{CahillNASA,CahillWaves},  and which displayed the sidereal effect, {\it viz} the earth time of the earth rotation phase was essentially fixed relative to sidereal time, i.e. the flow direction was fixed relative to the stars.
 
The zener diode detectors first used are known as  Random Number Generators (RNG) or Random Event Generators (REG). There are various designs available from manufacturers, and all claim that these devices manifest  hardware random quantum processes, as they involve the quantum to classical transition when a measurements, say, of the quantum tunnelling of electrons through a nanotechnology potential barrier, $\sim 10$nm thickness, is measured by a classical/macroscopic system.  According to the standard interpretation of the quantum theory, the collapse of the electron wave function to one side or the other of the barrier, after the tunnelling produces a component on each side, is purely a random event, internal to the quantum system.  However that interpretation had never been tested experimentally, until \cite{ZenerDiode}.   Data from  two REGs, located in Perth and London, was examined.   The above mentioned travel times were then observed.  The key features being a speed of $\sim$500km/s, and strong reverberation effects, see Fig.\ref{fig:Reverb}.  

This discovery revealed that the current fluctuations through a zener diode in reverse bias mode are not random, and data from collocated zener diodes showed almost identical fluctuations \cite{ZenerDiode}.  Consequently the zener diode detectors can easily be increased in sensitivity by using  zener diodes in parallel, with the sensitivity being proportional to the number of diodes used, see circuit diagram in \cite{ZenerDiode}.  That the quantum to classical transition, i.e. ``collapse of the wave function",  is induced by 3-space fluctuations, has deep implications for our understanding of quantum phenomena.

Using data from  REG's located in Perth and London, for Jan.1-3, 2013, and then doing a Fourier transform frequency analysis, we obtain the spectrum in the top two plots in Fig.\ref{fig:periods}.
The unfiltered power spectra from the two REGs show remarkable similarity to each other, and to the spectrum from the Frascati data. Again the dominant frequencies correspond to known earth vibration frequencies, \cite{Frequencies}, although there are long-period oscillations, common to all detectors, that are not known earth frequencies.

This new data shows that the time delays observed between Frascati and Rome  are to be expected, because of the strong reverberation effects seen in the zener diode detector data.  However the occurrence of the earth vibration frequencies is intriguing, and reveals new physics. Unlike the bar detectors it is impossible for any physical earth movement to  mechanically affect the zener diodes, and so all detectors are responding to dynamical space fluctuations caused by the oscillations of the matter forming the earth.  The key questions is what causes this ongoing activation of the earth modes? Are they caused by earthquakes or by the fractal 3-space waves exciting the earth modes?

\section{Conclusions}
The discovery of the quantum detection of gravitational waves, showing correlations between well separated locations, that permitted the absolute determination of the 3-space velocity of some 500km/s, in agreement with the speed and direction from a number of previous analyses, including in particular the NASA spacecraft Earth-flyby Doppler shift effect.  This discovery enables a very simple and cheap nanotechnology zener diode quantum gravitational wave detection technology, which will permit the study of various associated phenomena, such as solar flares, coronal mass ejections, earthquakes, eclipse effects, moon phase effects,  non-Poisson fluctuations in radioactivity \cite{Shnoll,ZenerShnoll}, and other rate processes, and variations in radioactive decay rates related to distance of the earth from the Sun, as the 3-space  fluctuations are enhanced by proximity to the sun.   As an example of these possibilities we have confirmed that the Amaldi {\it et al.} bar detectors did indeed detect gravitational wave events in 1980, but not of the form commonly expected, in particular gravitational waves do not travel at speed c, and there is no experimental or observational evidence supporting that claim.

\section{Acknowledgements}
This report is from the Flinders University Gravitational Wave Detector Project,   Australian Research Council Discovery Grant: {\it Development and Study of a New Theory of Gravity}.  Thanks  to  GCP  and its director Dr Roger Nelson for the availability of extensive and valuable data from the REG international network:  http://teilhard.global-mind.org/.  Giovanni B. Bongiovanni, Turin,  raised the earth vibration frequency  effect observed using a zener diode detector, and also confirmed the strong correlations between collocated detectors.

\end{document}